\documentclass{aa}  
\usepackage{caption}
\usepackage{graphicx}
\usepackage{txfonts}
\usepackage{color}
\usepackage{hyperref}
\usepackage{amsmath}
\usepackage{lineno}

\begin{document}

   \title{Into the thick of it: \\ ALMA 0.45~mm observations of HL Tau at 2~au resolution}


   \author{Osmar M. Guerra-Alvarado,
          \inst{1}
          Carlos Carrasco-Gonz\'{a}lez,
          \inst{2}
          Enrique Mac\'{i}as,
          \inst{3}
           Nienke van der Marel,
           \inst{1}
            Adrien Houge,
           \inst{3,4}
           Luke T. Maud,
           \inst{3}
           Paola Pinilla,
           \inst{5}
           Marion Villenave,
           \inst{6,7}
           Yoshiharu Asaki
           \inst{8,9,10}
           Elizabeth Humphreys
           \inst{3}
          }

   \institute{Leiden Observatory, Leiden University, PO Box 9513, 2300 RA Leiden, The Netherlands.
              \email{guerra@strw.leidenuniv.nl}
        \and Instituto de Radioastronom\'{i}a y Astrof\'{i}sica (IRyA), Universidad Nacional Aut\'{o}noma de M\'{e}xico (UNAM). Campus Morelia (Michoac\'an, M\'exico).
        \and ESO Garching, Karl-Schwarzschild-Str. 2, 85748, Garching bei Munchen, Germany.
        \and Department of Physics and Astronomy, University of Exeter, Exeter, EX4 4QL, UK.
        \and Mullard Space Science Laboratory, University College London,
3 Holmbury St Mary, Dorking, Surrey RH5 6NT, UK.
        \and Jet Propulsion Laboratory, California Institute of Technology, 4800 Oak Grove Drive, Pasadena, CA 91109, USA
        \and Dipartimento di Fisica, Università degli Studi di Milano, Via Giovanni Celoria 16, 20133 Milano, Italy
        \and Joint ALMA Observatory, Alonso de Córdova 3107, Vitacura, Santiago, 763 0355, Chile.
        \and National Astronomical Observatory of Japan, Los Abedules 3085 Oficina 701, Vitacura, Santiago, 763 0000, Chile.
        \and The Graduate University for Advanced Studies, SOKENDAI, Osawa 2-21-1, Mitaka, Tokyo 181-8588, Japan.
             }

   \date{}

 
  \abstract
   {}
   {To comprehend the efficiency of dust evolution within protoplanetary disks, it is crucial to conduct studies of these disks using high-resolution observations at multiple wavelengths with the Atacama Large Millimeter/submillimeter Array (ALMA).}
  { In this work, we present high-frequency ALMA observations of the HL Tau disk using its Band 9 centered at a wavelength of 0.45~mm. These observations achieve the highest angular resolution in a protoplanetary disk to date, 12 milliarcseconds (mas), allowing the study of the dust emission at scales of 2~au. We use these data to extend the previously published multi-wavelength analysis of the HL Tau disk, constraining the dust temperature, dust surface density, and maximum grain size throughout the disk. We perform this modeling for compact solid dust particles, as well as for porous particles.} {Our new 0.45~mm data traces mainly optically thick emission, providing a tight constraint to the dust temperature profile. We derive maximum particle sizes of $\sim1$ cm from the inner disk to $\sim60$ au. Beyond this radius, we find particles between 300 $\mu$m and 1 mm. The total dust mass of the disk is 2.1 M$_{J}$ with compact grains, and it increases to 6.3 M$_{J}$ assuming porous particles. Moreover, an intriguing asymmetry is observed at 32 au in the northeast inner part of the HL Tau disk at 0.45 mm. We propose that this asymmetry is the outcome of a combination of factors including the optically thick nature of the emission, the orientation of the disk, and a relatively large dust scale height of the grains preferentially traced at 0.45 mm. To validate this, we conducted a series of radiative transfer models using the RADMC-3D software. Our models, incorporating varying dust masses and scale heights, successfully replicate the observed asymmetry in the HL Tau disk. If this scenario is correct, our measured dust mass within 32 au would suggest a dust scale height H/R$>0.08$ for the inner disk. Finally, the unprecedented resolution allowed us to probe for the first time the dust emission down to a few au scales. We observed an increase in brightness temperature inside the estimated water snowline and speculate whether it could indicate the presence of a traffic jam effect in the inner disk.}
   {Our results showcase how 0.45 mm observations of protoplanetary disks can be used to robustly constrain their dust temperature radial profile. Additionally, the higher optical depths at this wavelength can be used to constrain the dust vertical scale height. Finally, these higher frequencies allow us to reach higher spatial resolutions that have the potential to resolve the region within the water snowline in disks.}

   \keywords{Planetary systems: Protoplanetary disk - Radio continuum: planetary systems
               }
\titlerunning{Into the thick of it}
\authorrunning{O. Guerra-Alvarado et. al.}
   \maketitle
%

\section{Introduction}

Over the past ten years, the Atacama Large Millimeter/submillimeter Array (ALMA) has made significant contributions to our understanding of protoplanetary disks by revealing substructures within them. These studies (e.g. \citealt{2015ApJ...808L...3A}, \citealt{2016PhRvL.117y1101I}, and \citealt{2018ApJ...869...17L}) have shown the frequent appearance of substructures in protoplanetary disks having different morphologies (\citealt{2013Sci...340.1199V}, \citealt{2016Sci...353.1519P}, and \citealt{2018ApJ...869L..43H}). These substructures play a significant role in protoplanetary disks allowing dust grains to accumulate and grow, facilitating the formation of planets by providing regions of higher dust density and trapping, and slowing down the radial drift of dust grains, which would otherwise hinder the planet formation process (\citealt{2012A&A...545A..81P}).

Most ALMA studies on protoplanetary disks are performed at wavelengths around 1~mm, where dust emission is relatively bright and high angular resolutions around 30 mas can be achieved. In some disks, similar studies have also been performed at longer wavelengths where the emission is known to be optically thinner, up to 3 mm with ALMA (e.g., \citealt{2021MNRAS.506.5117T}, \citealt{2018A&A...619A.161C},\citealt{2020ApJ...898...36L}) but also up to 1 cm with the Very Large Array (VLA) (e.g., \citealt{2016ApJ...821L..16C}, \citealt{2023AJ....166..186H}). These observations at long wavelengths allow us to study the properties of the dust grains albeit at the expense of angular resolution. Observations of protoplanetary disks with ALMA at shorter wavelengths, although providing the highest angular resolutions possible with ALMA ($\sim$10 mas), are usually avoided because of the higher atmospheric opacity and turbulence, leading to necessarily longer integration and calibration times, and the fact that dust emission is expected to be optically thick. However, there is also complementary information that can be extracted from studies at such wavelengths. For instance, optically thick emission impose better constraints on the temperature and the albedo of the dust particles, improving the results from multi-wavelength modeling of disks. Moreover, observations at short wavelengths are the only way to explore the dust content in the gaps whose emission is usually very optically thin to be detected at long wavelengths.

Recent multi-wavelength analyses of Class II protoplanetary disks (\citealt{2019ApJ...883...71C}, \citealt{2021A&A...648A..33M}, \citealt{2021ApJS..257...14S}, and \citealt{2022A&A...664A.137G}) have measured dust properties within these substructures. The dust sizes determined from these observations are around 1 mm. However, there is a discrepancy with measurements obtained from polarization observations (\citealt{2017ApJ...844L...5K}) which suggest dust sizes on the order of a few hundred micrometers. To reconcile these conflicting results, some studies by  \citealt{2017ApJ...844L...5K} and \citealt{2020ApJ...900...81O} have proposed the existence of two dust populations. The larger particles (mm-sized) would experience more efficient vertical settling, becoming decoupled from the gas and gathering towards the midplane of the disk.
On the other hand, smaller particles would remain mixed with the gas in higher layers of the disk (\citealt{2005A&A...443..185B}). More recently, \citealt{2023ApJ...953...96Z} proposed that large (sizes larger than 1 mm) and porous (porosities higher than 70$\%$) particles could reproduce both, multi-wavelength and polarization results. A very recent polarization multiwavelength study strongly supported the presence of porous particles in HL Tau \citep{2023arXiv230910055L}

Studies by \citealt{2016ApJ...816...25P}, \citealt{2020A&A...642A.164V}, \citealt{2021ApJ...912..164D}, \citealt{2022ApJ...930...11V} and \citealt{2023MNRAS.524.3184P} have focused on understanding the vertical structure of protoplanetary disks. Although limited in number, these studies suggest that Class II disks exhibit significantly low dust scale heights, particularly at longer wavelengths observed with ALMA. This is a consequence of the distribution of dust grains changing along the vertical direction due to settling (\citealt{2004A&A...421.1075D},\citealt{2021A&A...645A..70P}). Most constraints presented in these papers remain as upper limits (except in HD163296), with suggestions such as H(100au) < 1~au. Additionally, \citealt{2023MNRAS.524.3184P} discovered values < 4~au for certain DSHARP disks, particularly those with less favorable orientations. These findings were predominantly derived from the presence/depth of gaps and rings in the outer regions of the disks (outer 100 au) where the scale height in the inner regions remains mostly unknown. 

In this work, we present high-resolution Band 9 observations of HL Tau, a young stellar object in the Taurus-Auriga molecular cloud ($\sim 147$pc, \citealt{2018ApJ...859...33G}) with an estimated age of less than 1 million years (\citealt{2017A&A...607A..74L}). HL Tau is considered a Class I-II disk (\citealt{2008ApJS..176..184F}) and it was the first protoplanetary disk in which the presence of dark and bright rings was discovered (\citealt{2015ApJ...808L...3A}). Since it has been intensively observed in the last years with ALMA and VLA, high-quality data covering a wide range of wavelengths, between 0.9 mm and 1 cm, are currently available (\citealt{2016ApJ...821L..16C}, \citealt{2019ApJ...883...71C}). There have been also several previous studies modeling its continuum emission to infer the dust properties (e.g., \citealt{2016ApJ...818...76J}, \citealt{2016ApJ...816...25P}, \citealt{2017A&A...607A..74L}). From these, it has been well established that the dark rings are actually gaps in the dust distribution, i.e. regions of lower density than the adjacent bright rings. The origin of the substructures is still debated, but the most promising are planet-disk interactions (e.g., \citealt{2016ApJ...818...76J}, \citealt{2016MNRAS.459L...1D}).
Our new sensitive and high-resolution observations at 0.45~mm data are combined with the previous multi-wavelength observations to analyze the dust properties in the disk. The addition of shorter wavelength observations allowed us to improve previous modeling of the dust properties in the disk, as well as obtain new insights about the dust temperature very close to the protostar and the vertical structure of the disk.

\begin{figure*}[ht!]
\centering
\includegraphics[width=1.5\columnwidth,trim={0cm 0cm 0cm 0cm},clip]{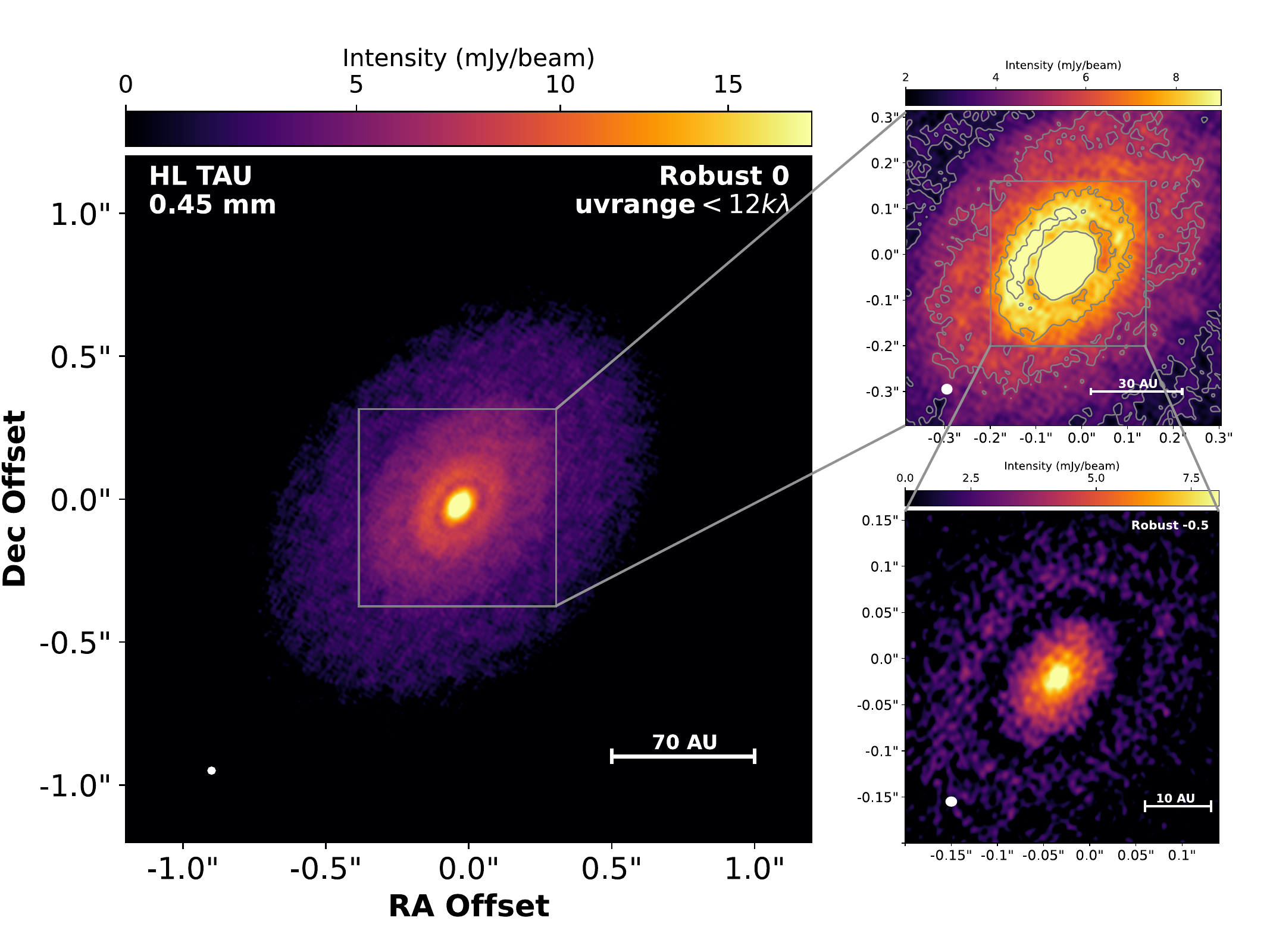}
\caption{ HL Tau Band 9 images. \textsc{Left panel}: Image using a robust parameter of 0 with a beam of 0.0226"$\times$ 0.021" \textsc{Right-top panel}: Zoom into the inner part of the HL Tau disk of the robust 0 images showing the asymmetry with gray contours at 11 times the rms and 33 times the rms. \textsc{Right-bottom panel}: Inner HL Tau disk from the robust -0.5 image with a beam of 0.0116"$\times$ 0.010".} 
\label{Fig1}
\end{figure*}

\begin{figure*}[ht!]
\centering

\includegraphics[width=1.5\columnwidth,trim={0cm, 0cm, 0cm, 0cm},clip]{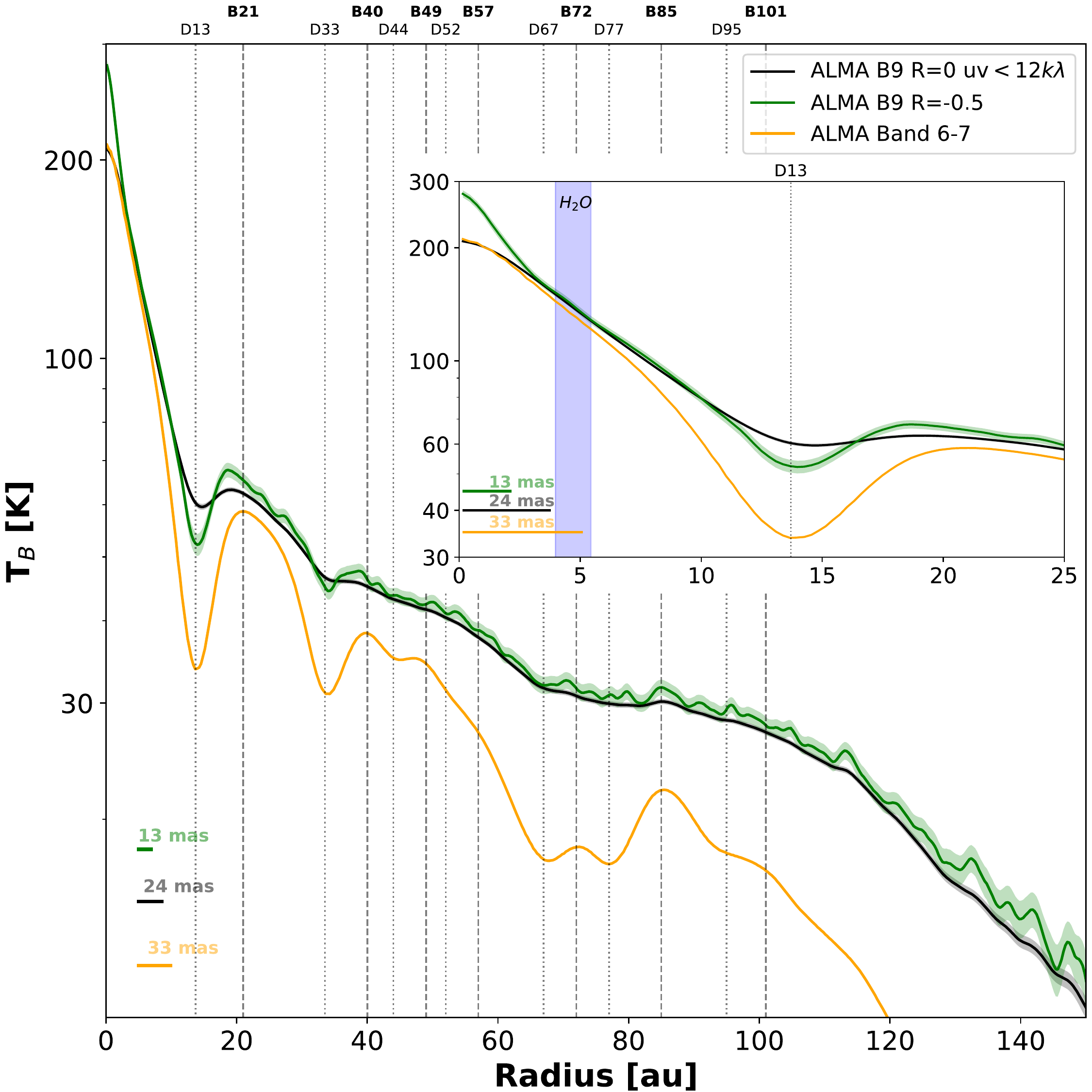}
\caption{Comparison of the HL Tau Band 6-7 (\citealt{2015ApJ...808L...3A}) and the HL Tau Band 9 radial profiles of the brightness temperature. In yellow we plotted the Band 6-7 ALMA image with the lowest resolution (33 mas), In black the Band 9 image with 22 mas, and in green the Band 9 image with 13 mas. The substructures are shallower in the radial brightness temperature for Band 9 compared to the Band 6-7 image, moreover, the emission from the Band 9 image extends up to 140 au. In the inside panel, we show the first 20 au of the HL Tau disk, we plotted the brightness temperatures of the three images and, assuming that T$_{disk}$=T$_{B}$ because of the optically thick regime, we displayed the position (in temperature) of the water snow line.} 
\label{Fig2}
\end{figure*}


\section{Observations}
We analyze archival observations obtained with the Atacama Large Millimeter/submillimeter Array (ALMA) at a wavelength of 0.45~mm. One of the archival datasets that we used was part of the test observations for the band-to-band calibration during the High-Frequency Long Baseline Campaign \citep{2020ApJS..247...23A}, project code: 2011.0.00005.E). The observations were taken during two execution blocks on November 3, 2017. The total observing time was 65 and 100 minutes for the first and second execution, respectively, which combined gives a 45-minute on-source time for HL Tau. The final data has 8 spectral windows, each of them covering a bandwidth of 2 GHz. We also used additional data at a lower resolution (project code: 2017.1.01178.S, PI: E. Humphreys) to complete the uv coverage at shorter baselines. These data have 8 spectral windows covering frequencies from 646-662 GHz with a bandwidth of 1.875 GHz each. The observations were taken from October 2, 2018 to October 18, 2018 with a total observing time on-source of 95 minutes.

We calibrated the data by using the calibration pipeline and scripts provided by the ALMA staff. We used version 5.6.1 of CASA (Common Astronomy Software Applications; \citealt{2007ASPC..376..127M} for self-calibration and cleaning in both data sets. We identified and flagged the spectral lines that were found in both data sets. We also averaged channels for the low and high-resolution data; the final self-calibrated datasets contain spectral windows of 8 channels of 200 MHz each.

We performed phase-only and amplitude self-calibration in both data sets. First, self-calibration was performed on the short baseline data. During the self-calibration process, we created the images using the task \textsc{clean} in CASA. The MTMFS deconvolver \citep{2011A&A...532A..71R} was used with scales of 0, 1, 3, and 5 times the beam size. We also used a Briggs weighting with a robust parameter of 0.5. For the low-resolution data set the self-calibration process resulted in a big improvement in the image quality. We applied 6 iterations of phase-only self-calibration decreasing the time solution interval down to 18s. Furthermore, we were also able to apply two rounds of amplitude and phase self-calibration (\texttt{calmode=ap}) until no improvement in the signal-to-noise ratio was found. By the end of the whole self-calibration process, we achieved an increase in the S/N by a factor of $\sim$17.4. We combined our short baseline data together with the long baseline data after aligning and correcting for proper motions and astrometric errors between our data sets. This was done by applying a Gaussian fit using CASA in the high-resolution image and using the tasks \textsc{fixvis} and \textsc{fixplanets} to apply shifts to the low-resolution data. After this, self-calibration was attempted on the combined dataset. We started with phase-only self-calibration, however, due to the low S/N at the long baselines, no improvement was found.

Lastly, self-calibration separately on the long-baseline data was also attempted by using the model obtained from the self-calibrated low-resolution data, but no improvement was found either. Since not enough solutions were found with any method to improve the quality of the combined data set, the self-calibration process was stopped. 

With our final data set at hand, we recalculated the weights using the task \textsc{statwt} in CASA. For the final images, we used the \textsc{tclean} task in CASA with the MTMFS deconvolver \citep{2011A&A...532A..71R} using nterms=1 and scales at 0, 1, 3, and 5 times the beam size. We used a pixel size 10 times smaller than the beam size for all of the images. We explored different weightings: natural, uniform, and Briggs with different robust values. Finally, two images are used for this study. The first one, made by setting the parameter robust to 0 and using baselines shorter than 12 $k\lambda$ in length, gives us a good compromise between resolution and sensitivity and was used for the study of the large scale emission of the disk. This image has an rms noise of 270~$\mu$Jy~beam$^{-1}$ and a beam size of $\sim$23 mas. A second image was done by setting the parameter robust to -0.5, which results in a beam size of $\sim$12 mas and an rms noise of 380~$\mu$Jy~beam$^{-1}$. This higher angular resolution image allowed us a detailed study of the innermost part of the disk ($\lesssim$20~au). Both images are shown in Figure \ref{Fig1}.

In order to study the radial properties of the disk, we also obtained radial profiles of the emission and convolved the images to a round beam (robust 0, 24 mas and robust -0.5, 13 mas). For this, we used the known inclination and position angles of the disk, 46.72$^{\circ}$ and 138.02$^{\circ}$, respectively \citep{2015ApJ...808L...3A}. However, due to the inclination of the disk, there is a loss in angular resolution in the NE-SW direction. Since the S/N of the image is very high, we could avoid this loss in resolution by simply averaging emission azimuthally within $\pm$0.2~ rad of the major axis of the disk. Radial profiles of the brightness temperature obtained from the intensity profiles of both images are shown in Figure \ref{Fig2}.

\section{Results}

In Figures \ref{Fig1} and \ref{Fig2}, it can be seen that our images at 0.45~mm recover emission from the whole disk with a radius of $\sim$1$\arcsec$ or $\sim$140~au, similar to previous ALMA images at longer wavelengths (see e.g., \citealt{2019ApJ...883...71C}, \citealt{2015ApJ...808L...3A}). Our 0.45~mm image also shows some substructures but, as expected, due to the higher optical depth at this wavelength, substructures are detected with a lower contrast between dark and bright rings. This suggests that the gaps contain small particles traced by the 0.45~mm image, these particles are more coupled with the gas and might fill the gaps more than bigger particles observed at longer wavelengths (\citealt{2012A&A...538A.114P},\citealt{2013A&A...560A.111D}). Furthermore, we observed a slight deviation in the positions of the gaps and rings at 0.45~mm when compared to the radial profiles at other wavelengths (see Figure \ref{Fig2}), which suggests that there might be slight variations in the depth and width of the disk substructures at different wavelengths.

In the right-top of Figure \ref{Fig1}, we show a zoomed-in view of the inner 50~au. An asymmetry can be seen in the 0.45~mm data that is not seen at longer wavelengths: the NE part of the first ring is more intense than the SW part of the ring. 

In the right-bottom panel of Figure \ref{Fig1}, we show our highest angular resolution image which allowed us to resolve, for the first time, the most internal part of the disk ($\lesssim 20$~au). In this image, we can observe the first ring and a very deep gap. 

This first gap, centered at 13~au, while it shows some emission in the low resolution 0.45~mm, it appears very dark in our highest angular resolution image (see Figures \ref{Fig1} and \ref{Fig2}). 

These observations provide a clearer view of the gap's structure, revealing that it is likely deeper and more depleted of dust particles than what was initially observed with lower resolution at 0.87 mm and 1.3 mm. Moreover, the fact that we also detect a very deep gap in the 0.45~mm image strongly suggests that this gap is actually devoid of small dust particles as well.

In the inner panel of Figure \ref{Fig2}, it can be seen that the brightness temperature increases within 2.5~au when observed at the highest angular resolution. This could potentially be attributed to substructures within the inner disk that are becoming resolved for the first time thanks to the higher angular resolution of our 0.45~mm data.

In the following section, we first discuss the updated modeling of the radial dust properties using all the currently available data of HL Tau, then the origin of the new features found in the 0.45~mm images, the asymmetry, and the substructure within the inner disk.

\section{Discussion}

\subsection{Modeling of the dust properties}\label{sec4.1}

\begin{figure*}[ht!]
\centering
\includegraphics[width=1.5\columnwidth,trim={0cm, 0cm, 0cm, 0cm},clip]{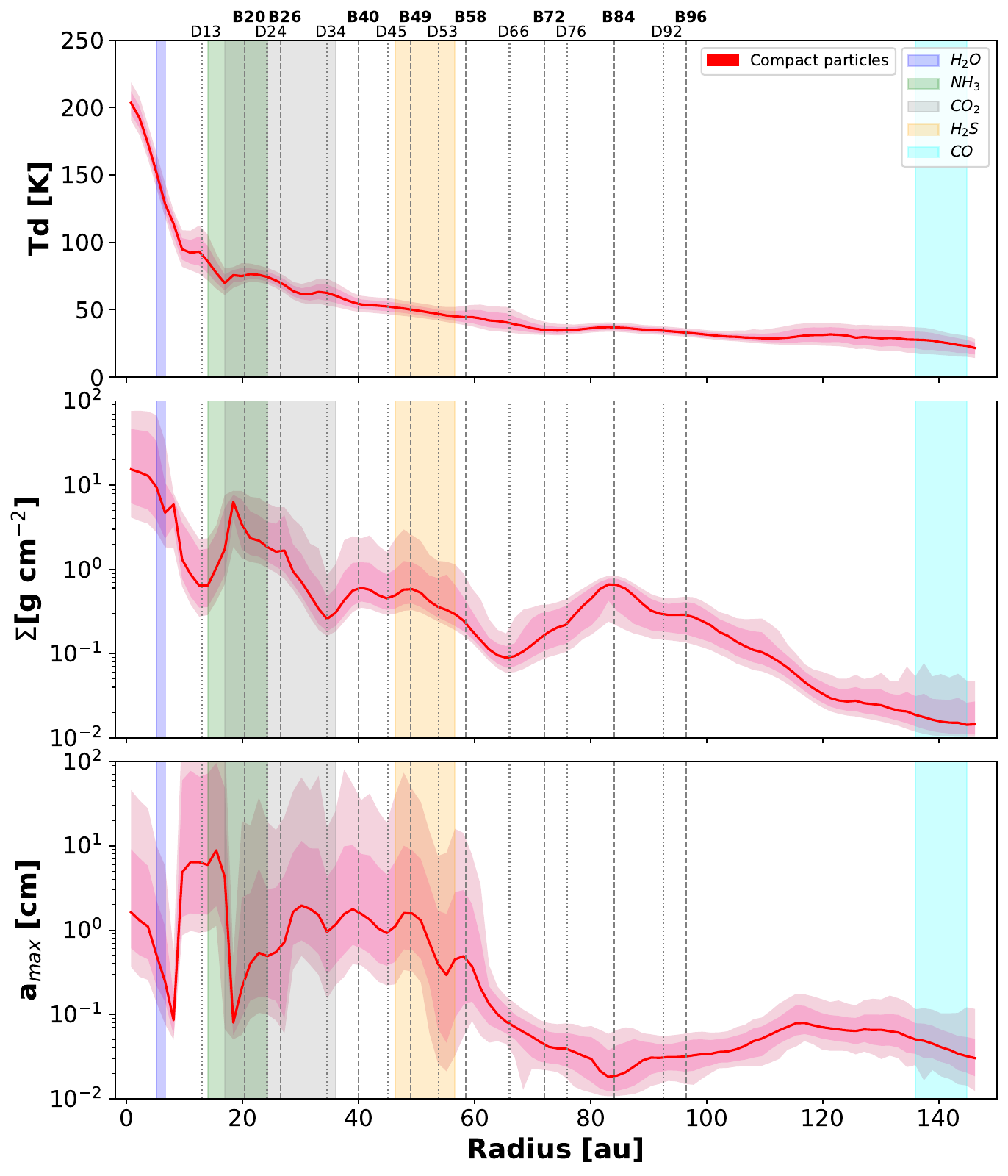}
\caption{Dust parameters at each radius using compact dust particle opacities, the position of the most common snowlines are plotted in colors. The vertical dotted and dashed lines are the positions of the gaps and rings, respectively, as seen from the surface density parameter.  \textit{Top panel}: Dust temperature at each radius, the temperature is very well constrained due to the optically thick nature of the Band 9 image. \textit{Middle panel}: Dust surface density at each radius. As expected the density increases in the bright rings and decreases in the gaps.  \textit{Bottom panel}: Maximum particle size at each radius, the particle sizes are around 1 mm - 1 cm. } 
\label{Fig3}
\end{figure*}

\begin{figure*}[ht!]
\centering
\includegraphics[width=1.5\columnwidth,trim={0cm, 0cm, 0cm, 0cm},clip]{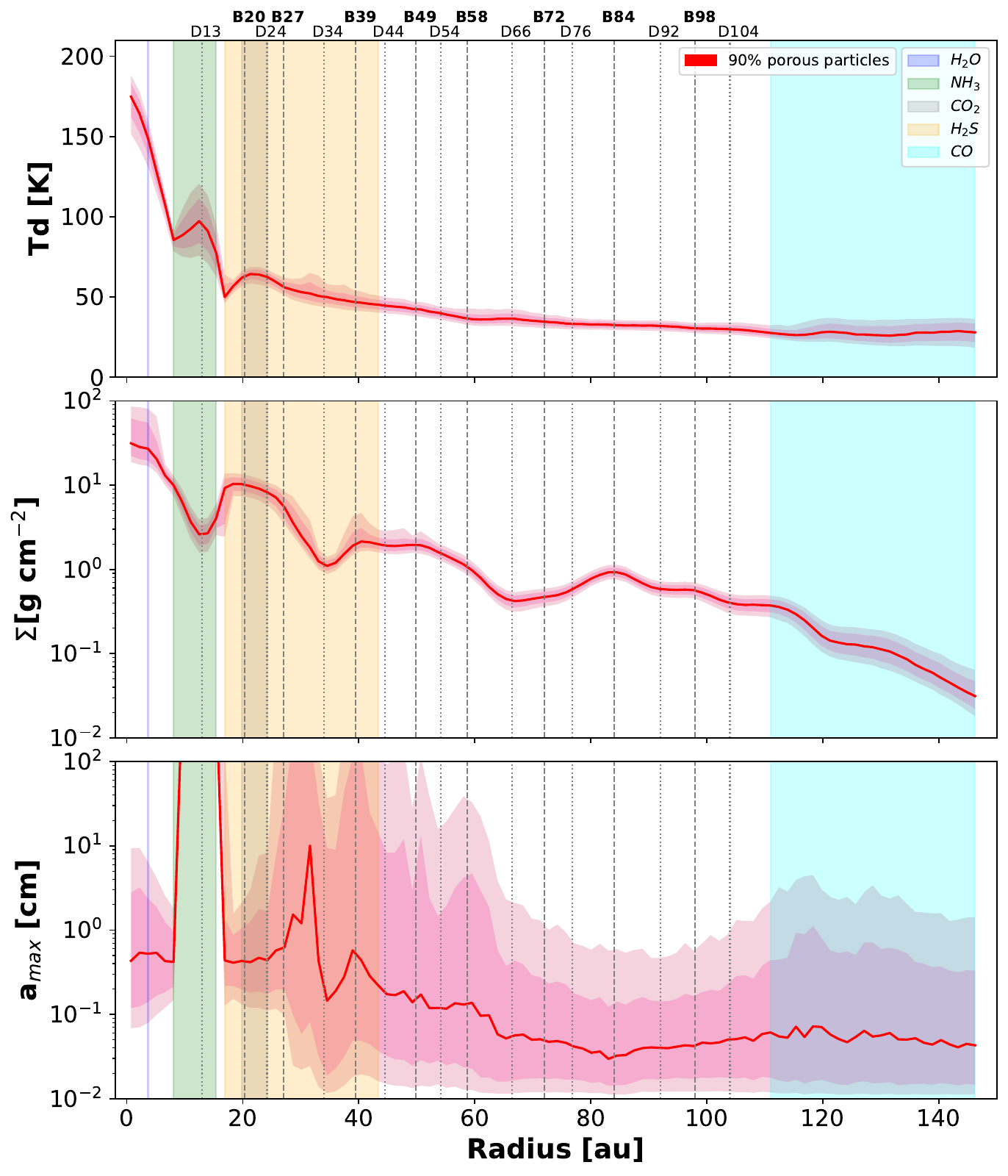}
\caption{Same as Figure \ref{Fig5} but using a 90\% porosity for the dust particles and for the dust opacities. The dust temperature is lower, the surface density is higher around the gaps and the maximum particle sizes are similar, however, at the 13 au gap there is a huge increase in the particle size. } 
\label{Fig4}
\end{figure*}

Previous to this work, there have been several studies modeling the dust properties in the HL Tau using the available data at that moment (e.g., \citealt{2016ApJ...816...25P}, \citealt{2017A&A...607A..74L}, \citealt{2019ApJ...883...71C}). We perform a new modeling procedure by including the new 0.45~mm image reported here which will provide valuable constraints on the dust temperature and will enhance the comprehensiveness of our multi-wavelength analysis. Thus, this new analysis includes images at 0.45, 0.87, 1.3, 2.1 and 7.0 mm. We created a new 0.45 mm image with a robust parameter of 1.0, using baselines < 12 $k\lambda$ in length as previously described for the robust 0.0 image. Then, we convolved it with a 50 mas Gaussian beam and extracted radial profiles in wedges along the major axis within radial beams of approximately 0.01 arcseconds. We also extracted radial profiles at the other wavelengths applying the same procedure to the images presented in \citet{2019ApJ...883...71C}. Subsequently, each point in the radial profiles was modeled independently. We note that each point in the model is not independent from its neighboring points due to the effects of the beam. Consequently, the contrast of the model parameters between rings and gaps will be slightly smoothed in our results. We follow a similar approach to \citep{2019ApJ...883...71C}, i.e., at each radius, we fit the spectral energy distribution (SED) in the millimeter wavelength range using a 1D slab model given by:
\begin{eqnarray}
I_{\nu}=B_{\nu}(T)[(1-\exp( \tau_{\nu}/\mu))+\omega_{\nu}F(\tau_{\nu},\omega_{\nu})],
  \label{eq:emergent intensity (Sierra et al. 2019)}
 \end{eqnarray}
where, 
 \begin{eqnarray}
F(\tau_{\nu},\omega_{\nu})=\frac{1}{\exp(-\sqrt3\epsilon_{\nu}\tau_{\nu})(\epsilon_{\nu}-1)-(\epsilon_{\nu}+1)}\textsc{x}\nonumber\\ \left [  \frac{1-\exp(-(\sqrt{3}\epsilon_{\nu}+1/\mu)\tau_{\nu})}{\sqrt{3}\epsilon_{\nu}\mu+1}+\frac{\exp(-\tau_{\nu}/\mu)-\exp(-\sqrt{3}\epsilon_{\nu}\tau_{\nu})}{\sqrt{3}\epsilon_{\nu}\mu-1}  \right ],
  \label{eq:F(tau,omega) (Sierra et al. 2019)}
 \end{eqnarray}

\noindent 
where  $\omega_{\nu}=\frac{\sigma_{\nu}}{\kappa_{\nu}+\sigma_{\nu}}$ is the albedo, which is defined by the scattering coefficient ($\sigma_{\nu}$) and the absorption coefficient ($\kappa_{\nu}$), $\tau_{\nu}$ = $\Sigma_{dust}\chi_{\nu}$, where $\chi_{\nu}= {\kappa_{\nu}+\sigma_{\nu}}$, $\mu=cos(i)$ considers the inclination effects by correcting the optical depth by this inclination ($i$), and $\epsilon_{\nu}=\sqrt{1-\omega_{\nu}}$. This model assumes that most of the dust content is settled in the mid-plane (see \citealp{2019ApJ...876....7S} and \citealp{2019ApJ...883...71C} for details). While this assumption holds well for the outer HL Tau disk, it is not met within the initial $\sim$ 32 au. A particle size distribution that follows a power-law distribution was also assumed, specifically, $n(a) \propto a^{-p}$. We set the value of the exponent $p$ to 3.5, which is a common value used in the Interstellar Medium (ISM) \citep{1977ApJ...217..425M}. 

The dust opacity values were obtained assuming the DSHARP dust composition and following the effective medium approximation \citep{2018ApJ...869L..45B}. As usual in the study of protoplanetary disks, we first consider compact solid spherical particles. However, we also perform an analysis by considering spherical porous dust particles with a porosity of 90\%. This is motivated by very recent results which strongly suggest that the HL Tau disk contains porous dust particles (\citealt{2023ApJ...953...96Z} and \citealt{2023arXiv230910055L}). 

The opacity tables for the porous particles were obtained using the same procedure as \citep{2018ApJ...869L..45B}: the grain is assumed to be composed of small monomers within a void matrix. The optical properties of the monomers are computed first using the Bruggeman rule. This same rule is also used for the compact grains. The monomers are then mixed with a void matrix using the Maxwell-Garnett rule. It is well-established that the fluxes and, consequently, dust masses exhibit significant dependence on the assumed dust composition, particularly the quantity of carbon. For instance, when using the opacities presented by \citealt{2010A&A...512A..15R} and \citealt{2022A&A...668A.104S} the fluxes they inferred were higher which made the spectral indices of typical disks more easily explained. Therefore, the composition of the grains plays a crucial role, and the total dust mass derived in the subsequent section should be approached with caution.

We model the SED of the disk at each radius independently and we fitted the dust parameters (T$_{dust}$, a$_{max}$ and $\Sigma_{dust}$). We assume a flux calibration uncertainty of 10\% at 0.45~mm, 0.9 mm, 1.3 mm and 7 mm, and 5\% at 2.1 mm. We compute the posterior probability distribution of our model at each radius using the Markov Chain Monte Carlo (MCMC) implementation within \texttt{emcee} \citep{2013PASP..125..306F}. We assume uniform priors in dust surface density and maximum grain size. Following \citealt{2021A&A...648A..33M}, we use a conservative (i.e., wide) prior for the dust temperature based on the expected radial temperature profile for a passively irradiated disk, $T_d = (\phi {L}_{\star}/8 \pi r^2 \sigma_{\rm SB})^{0.25}$, assuming ${L}_{star} = 6 \pm 5~{L}_{\odot}$ (i.e., a very conservative uncertainty on the stellar luminosity) and a range in $\phi$ (i.e., flaring angle) between 0.005 and 0.3. This prior is wide enough to have no effect on the posterior of the temperature at most radii, only helping avoid reaching unphysically high dust temperatures at a few radii within the wide gap between 60 and 90 au, where the emission is optically thin and, therefore, the dust temperature and dust surface density are highly degenerated.

In Figures \ref{Fig3} and \ref{Fig4}, we show the dust parameters for the HL Tau disk obtained from the fitting with the two different particles, solid and porous, respectively. In Appendix A, we show the model radial intensity profiles at each wavelength and their uncertainties, obtained from the median, 16th, and 86th percentiles of 500 random SEDs computed from the MCMC chains. As anticipated, the dust temperature is very well constrained for both cases, compact and porous particles. We also note that the surface density is well-constrained in both cases. Moreover, our findings reveal that dust particles in the HL Tau disk are relatively large, ranging from millimeters to centimeters in size, these resemble well with models of dust evolution with traps located at different distances \citep{2015A&A...573A...9P}.

Our models find a sudden, and likely unphysical, increase in particle size at the 13 au gap. However, figures \ref{Fig:A1} and \ref{Fig:A2} show that our models are in fact incapable of reproducing the observed emission at these radii. We believe the reason for this is our limited angular resolution, paired with the fact that the 13 au gap is highly depleted of large dust particles. As such, the SED at high frequencies is dominated by the small grains present in the gap, while at longer wavelengths the emission comes mostly from the beam smearing of the emission at the 20 au ring. Higher angular resolution observations at long wavelengths, only feasible with the upcoming ngVLA, are therefore necessary to accurately constrain the dust properties at these radii.

A distinct gap at 70 au, wide and with almost no large particles, is evident in \ref{Fig3} and \ref{Fig4}, consistent with prior observations by \citealt{2019ApJ...883...71C}. Interestingly, an increase in dust surface density is then present at the 84 au and 96 au rings, with grain sizes still remaining at a few hundreds of microns. We note that an infalling streamer has been found to be interacting with the HL Tau disk at these radii \citep{2022A&A...658A.104G}. We speculate that the small dust particles of this late infalling material might be responsible for the creation of this dust ring, which would explain why the particles at these radii have not yet grown up to mm/cm sizes. More detailed modeling of the interaction between the disk and infalling streamers would be needed to confirm this scenario.

Furthermore, clear distinctions emerge between the compact and porous particle solutions. In general, the model using compact particles exhibits lower temperatures and dust surface densities across the entire range of radii. The particle sizes predicted by the porous dust composition appear to be slightly smaller, although we note that they have a significantly wide uncertainty range. 

Given that our new 0.45~mm data have allowed us to better constrain the radial dust temperature profile, we can more robustly compare the position of the rings and gaps in HL Tau with the expected positions of the snowlines of some of the most relevant volatiles in protoplanetary disks. We plot in Figures \ref{Fig3} and \ref{Fig4} the expected radii of the snowlines of H$_2$O, NH$_3$, CO$_2$, H$_2$S, and CO. Except for the H$_2$O snowline, the uncertainties in the sublimation temperatures of these volatiles now dominate the uncertainty in their positions. These large uncertainties still make it hard to extract any robust conclusion, despite our improved temperature profile. Given the abundance of substructures in HL Tau, these snowlines will always fall close to a ring or gap. This makes it hard to confidently reject that snowlines are at the origin of at least a few of the ring substructures in HL Tau, as previously proposed (\citealp{2015ApJ...806L...7Z}; \citealp{2019ApJ...883...71C}). However, many other substructures appear to be far from the expected snowlines as in other protoplanetary disks \citep{2023ASPC..534..423B}. This applies mostly to the outer substructures, between 65 and 98 au, but also some between 36 to 45 au. We therefore conclude that mechanisms other than snowlines (e.g., MHD effects or planet-disk interactions) must be responsible for most, if not all, the ring substructures in HL Tau. 

We infer a total dust masses of 2.1$^{+1.88}_{-0.84}$ and 6.3$^{+1.57}_{-1.04}$ M$_{J}$ from the compact and porous particles, respectively. The dust masses assuming porous particles are thus larger by a factor of $\sim3$, as previously reported by \citealt{2023ApJ...953...96Z}. Our measurement of the dust mass with compact grains is two times higher than the mass obtained in \citealt{2019ApJ...883...71C} assuming compact particles with the same composition as ours (1.04 M$_{J}$). The primary difference between their analysis and ours is the inclusion of Band 9 in our sample.  
These additional data have allowed us to more robustly constrain the dust temperature, which in turn affects the rest of the modeling. 
We obtained lower dust temperatures across all radii. The surface density is comparable at the inner radius but it has a more rapid decline compared to their study. Additionally, our analysis reveals more pronounced substructures in the surface density. Notably, we observe larger dust particle sizes by approximately one order of magnitude within 60 AU, but beyond this point, the particle sizes decrease more significantly compared to the findings in \citealt{2019ApJ...883...71C}. These larger particles inside 60 au have lower absorption opacities, which explains why we obtain a higher dust mass than \citealt{2019ApJ...883...71C}.
Additionally, \citealt{2020MNRAS.493L.108B} estimated a lower limit for the total gas mass of the disk to be 209.5 M$_{J}$. This reconciles well with our results of the compact dust particles if we consider a gas-to-dust ratio of a 100. However, the elevated mass associated with porous dust particles from our analysis might imply a reduced gas-to-dust ratio $\sim$ 33 when compared to the total disk mass of 209.5 M$_{J}$. 
This suggests that the total gas mass of the disk might indeed still be larger than 209.5 M$_{J}$.

We note that our assumption of a 1-D slab model might not be accurate in the inner radius < 32 AU, where the emission at 0.45 mm is so optically thick that it could be tracing layers of the disk above the midplane (see section \ref{sec4.2}). If all wavelengths are effectively probing adjacent and close layers in the disk, indicating minimal variations in temperature and dust size distribution between them, then the assumption could still be valid. The midplane of protoplanetary disks is generally vertically isothermal, so it is most likely that, if our observations are tracing significantly different layers of the disk, this has only an effect on the maximum grain sizes traced at each wavelength. \citet{2020ApJ...892..136S} explored these effects in more detail and found that the high optical depths in the innermost regions of disks could result in incorrectly measuring the maximum grain size to be of a few hundreds of microns. While we cannot discard that our modeling is affected by these effects, the fact that we find maximum grain sizes of $\sim5$ mm in the inner region, instead of the 200-300 microns found by \citet{2020ApJ...892..136S} suggests that our modeling is not severely affected by the high optical depths.

\subsection{Azimuthally asymmetric emission in the inner disk}\label{sec4.2}

\begin{figure}[ht!]
\centering
\includegraphics[width=0.5\textwidth,trim={0cm 0cm 0cm 0cm},clip]{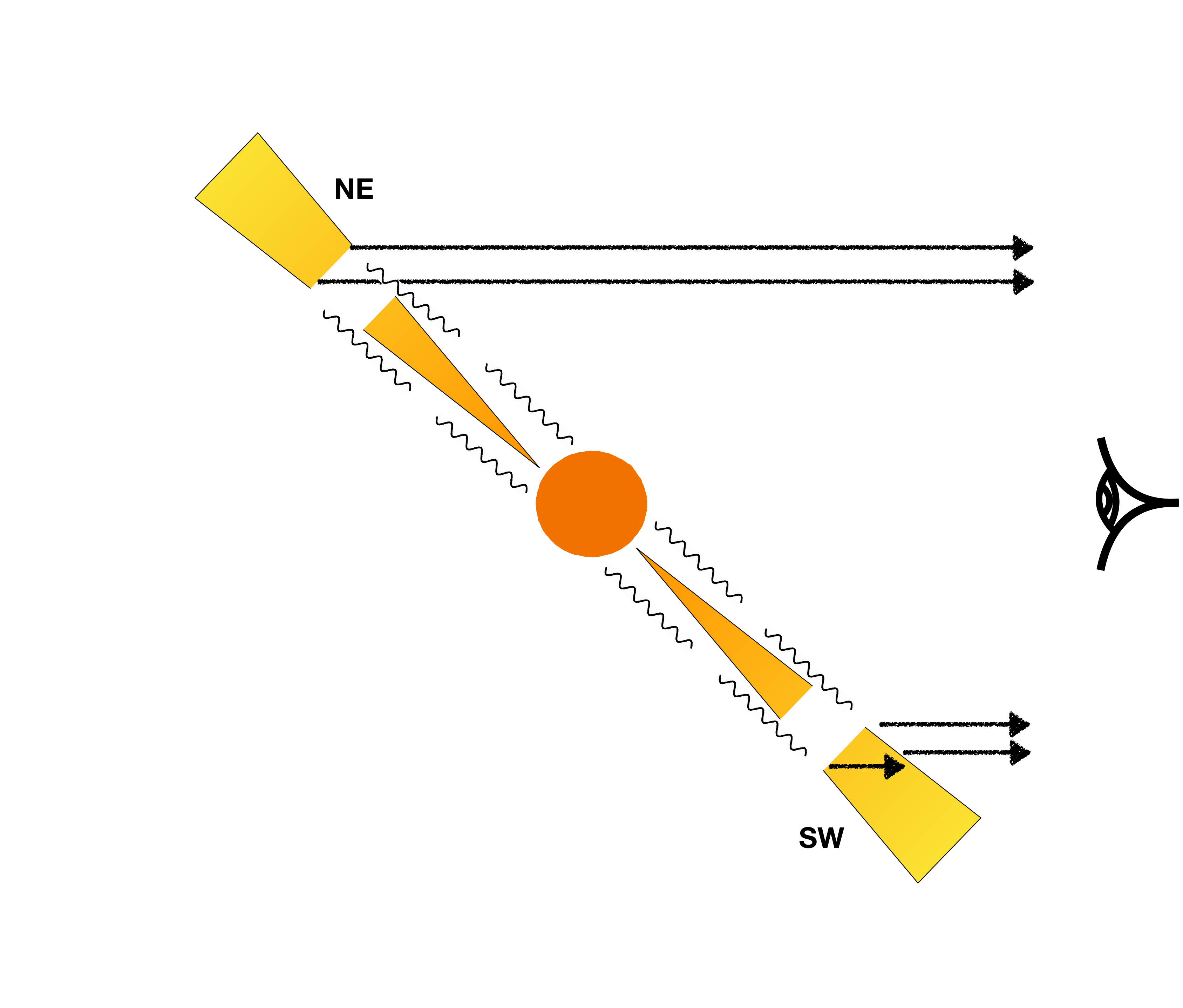}
\captionof{figure}{Diagram explaining the scenario proposed for the origin of the asymmetry seen in HL Tau at 0.45~mm.} 
\label{Fig5}
\end{figure}

As mentioned in Section 3, the first ring, centered at 21 au, shows an azimuthal asymmetry at 0.45~mm that becomes fainter at longer wavelengths and we believe this is contingent upon the resolution and sensitivity of the observations. This asymmetry appears as an increase in brightness at the northeastern side of the ring, covering almost 180 degrees and centered on the minor axis of the disk (see Fig. \ref{Fig1}). The position of this asymmetry strongly suggests that it is caused by the combination of the disk's inclination and its high optical depth at the observed wavelength (see Fig. \ref{Fig5}). Specifically, in the northeastern (NE) part of the disk, we are directly observing the emission from the internal wall of the first ring, which is directly irradiated by the central star, increasing its temperature. 
On the other hand, in the southwestern (SW) part of the disk there is no direct line of sight toward the wall, so the wall emission is blocked by the optically thick material in the line of sight (see Figure \ref{Fig5}). 

Given that the HL Tau outflow is blue-shifted in the NE and red-shifted in the SW \citep{2014AJ....147...72L}, we can infer that the NE corresponds to the far side of the disk, while the SW corresponds to the close side of the disk to the observed, this explains the position of the asymmetry in that side of the minor axis of the disk.

The asymmetry, in this case, is fundamentally similar to those found in Class 0/I objects (e.g \citealt{2024A&A...681A..82G},\citealt{2023ApJ...951....9L},\citealt{2023ApJ...951....8O}) where the dust particles did not have time to settle or are affected by significant turbulence preventing the grains to fall into the midplane (larger dust scale height), remaining optically thick even at longer wavelengths with ALMA. On the other hand, this asymmetry is fundamentally different from the large-scale asymmetries observed in other disks, where the effects become more pronounced at longer wavelengths (e.g., IRS48 \citep{2015ApJ...810L...7V} and HD142527 \citep{2015ApJ...812..126C} which is contrary to the observed behavior in HL Tau. These asymmetries are commonly associated with dust traps where dust can grow up to cm sizes.

In order to confirm that the asymmetry is caused by a geometric effect, we use a toy model to reproduce the observed asymmetry. We use RADMC-3D \citep{2012ascl.soft02015D} to perform radiative transfer models at 0.45~mm while varying parameters such as the disk dust mass and scale height. We then compare the model emission with the observed data to gain insights into the disk's physical properties and vertical structure. In this specific radiative modeling, a generic protoplanetary disk model was utilized within RADMC-3D. The disk dust masses and scale height were adjusted to explore different scenarios and to reproduce the observed asymmetry. The \texttt{optool} package \citep{2021ascl.soft04010D} was employed to compute the DSHARP dust particle opacities using a$_{min}$ as 0.050 $\mu$m and a$_{max}$ as 3mm. For this toy model, we are only interested in the inner regions of the disk, so we consider only emission up to a radius of 32~au (the location of the second gap). We assumed that all the grains are at the same height and we set a small inner radius $R_{in}$=1~au. A gap between 8~au and 17~au was included in the model to capture the effects of the asymmetry at the outer edge of the gap. 

The generic protoplanetary disk model uses a density distribution that goes as:

 \begin{eqnarray}
 \rho(r,z) =\frac{\Sigma(r)}{H_{p}\sqrt{2\pi}}\exp(-\frac{z^{2}}{2H^{2}_{p}}),
 \label{eq: Density distribution}
 \end{eqnarray}

\noindent where $r$ is the distance to the star from the disk, $H_{p}$(r) is the dust scale height of the disk and $\Sigma(r)$, is the dust surface density defined as:

 \begin{eqnarray}
\Sigma(r)= \Sigma_{0}(\frac{r}{r_{out}})^{-1},
 \label{eq: Surface Density distribution}
 \end{eqnarray}

The dust scale height in the radiative modeling is defined as a power-law-like dependence given by:

 \begin{eqnarray}
H_{p}(r)=H_{100}(\frac{r}{100 AU})^{1+\Psi},
 \label{eq: Scale Height}
 \end{eqnarray}
 
\noindent where $\Psi$ is the value of the flaring index set by default to 0.14 and $H_{100}$, is the value of the scale height at a distance of 100 AU from the central star. 

Finally, a total stellar mass of 1.7 M${\odot}$ \citep{2016ApJ...816...25P}, a stellar luminosity of 6 $ L_{\odot}$ and an effective temperature of 4395~$K$ \citep{2004ApJ...616..998W} were assumed. In addition, the inclination angle for HL Tau was set to 46.72$^{\circ}$ and the distance to 147 pc.\\

Several models were run by varying the dust scale height and the dust disk mass within 32~au (M$_{Disk}^{< 32 au}$) based on previous studies (\citealt{2016ApJ...816...25P}; \citealt{2017A&A...607A..74L}). We showed this parameters and their combination in Table \ref{table:RADMC-3D parameters}. 
\begin{table}[h!]
\caption{Model parameters explored}
\centering
\begin{tabular}{ccccc}
\hline \hline
M$_{Disk}^{< 32 au}$ & 1 M$_{J}$ & 3 M$_{J}$ & 5 M$_{J}$  \\

\hline

H/R & 0.01  & 0.01  & 0.01   \\ 
& 0.03  & 0.03 & 0.03 \\
 & 0.05 & 0.05 & 0.05 \\
 & 0.08 & 0.08 & 0.08\\
 & 0.1 & 0.1& 0.1 \\

\hline
 \label{table:RADMC-3D parameters}
\end{tabular}
\end{table}

\begin{figure*}[ht!]
\centering

\includegraphics[width=520pt,trim={0cm 0cm 0cm 0
cm},clip]{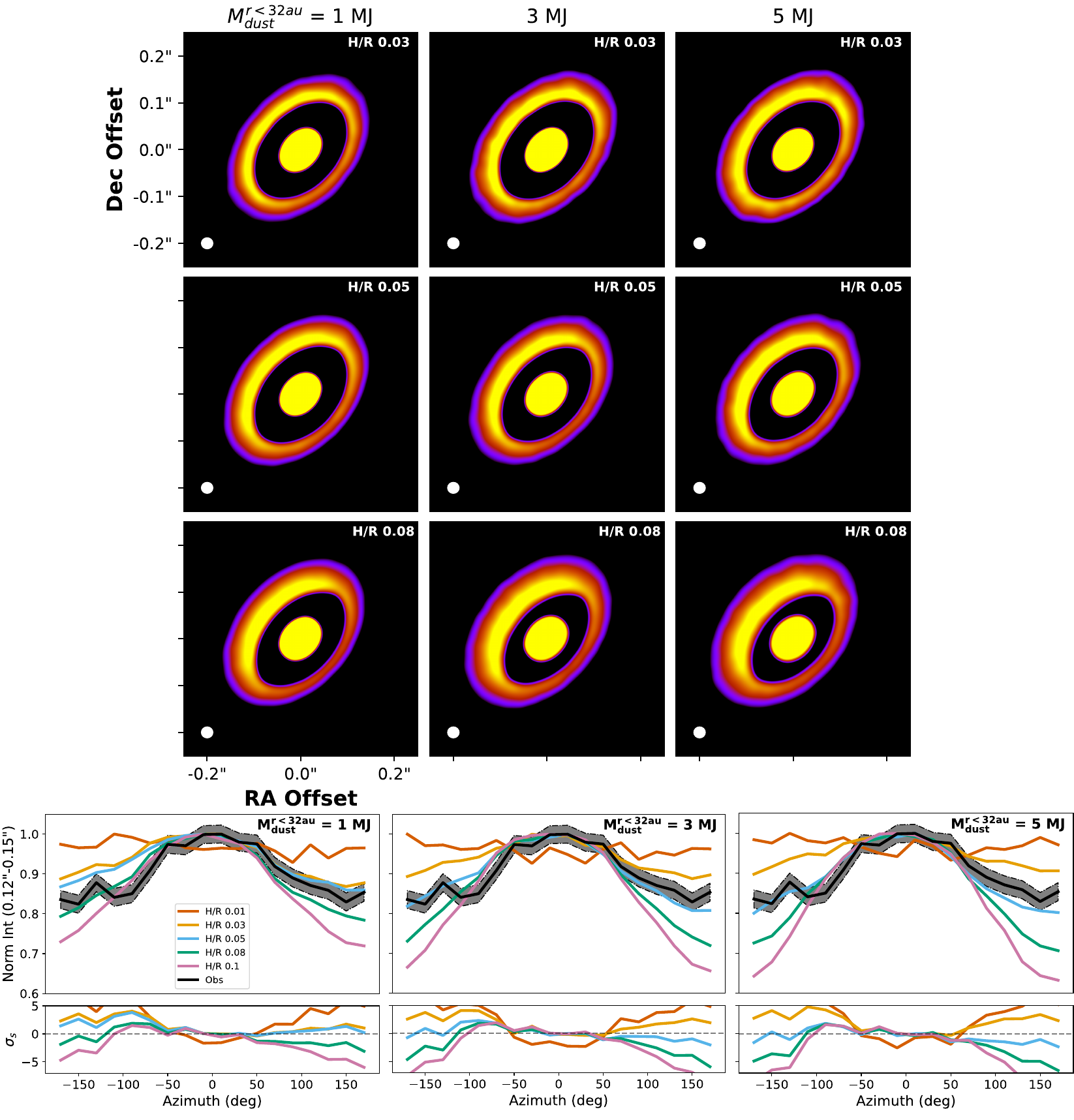}
\caption{RADMC-3D models of the inner HL Tau disk compared with the Band 9 observation. Top Figure: Convolved images of the 9 different models with RADMC-3D that better reproduce the observation, corresponding to masses of 1 M$_J$, 3 M$_J$, and 5 M$_J$, paired with scale height values of 0.08, 0.05, and 0.03 $H_{p}$. The disk mass is defined as the dust mass within 32 au. Bottom panels: Normalized azimuthal profiles considering the emission between 0.12 and 0.15 arcseconds in the HL Tau Band 9 observation and all the RADMC-3D models. The lowermost panels depict the difference, measured in sigmas, between the observation and the respective models}  \label{Fig6}
\end{figure*}

The model images were then convolved using a Gaussian beam with the same beam as the 0.45~mm image. The final convolved model images that seem to better reproduce the emission of the observation are shown in Figure \ref{Fig6}. We also extracted the deprojected emission from each pixel in a ring ranging from 0.12$\arcsec$ to 0.15$\arcsec$ and around 360$^{\circ}$ to capture the increased emission in the asymmetry. We show the normalized azimuthal profiles from the observation and all models in Figure \ref{Fig6}. We note that our goal with this toy models is to show that this type of asymmetries can be obtained at 0.45~mm and that one could constrain the dust mass and scale height from these images. As such, we do not try to reproduce the exact flux of our observations, since that would require a much more detailed modeling varying more disk parameters. The azimuthal profiles for 1, 3, and 5 Jupiter masses (Inner disk mass + ring mass) are also shown in Figure 6.

Our modeling shows that an asymmetry like the observed one in HL Tau can be produced by the combined effects of disk orientation, high optical depth, and larger scale heights at 0.45~mm.
From Figure \ref{Fig6}, it is evident that the dust mass plays a significant role in reproducing the observed asymmetry at 0.45~mm and in models with not a large enough scale height, the emission doesn't produce an asymmetry.

If the asymmetry arises from the scenario we outlined before, we find that a disk dust mass (within 32~au) of approximately 1 $M_J$, combined with a scale height H/R of 0.08, a disk dust mass of 3 $M_J$ with a scale height H/R of 0.05, or a disk dust mass of 5 $M_J$ with a scale height H/R slightly smaller than 0.05 best reproduces the observed asymmetry in the observations. In other words, the higher the dust mass, the less dust scale height one needs to reproduce an asymmetry. Because of this, obtaining an independent estimation of the dust mass within 32~au would allow for a better constraint on the scale height of the disk. We employ the mass obtained from our multi-wavelength analysis in section \ref{sec4.1}, acknowledging its known limitations, to give an initial approximation for the scale height in this specific region of the disk. From this analysis, we obtained a disk dust mass inside the first 32 au of 0.86$^{+1}_{-0.368}$ M$_J$.  Therefore, if our proposed scenario is the reason of the asymmetry, our derived mass would imply a scale height $\sim$>= 0.08 H/R in that specific region of the disk. As seen in Figure 6, to more effectively determine the scale height in subsequent studies we need a more restrictive constraint on the disk dust mass together with a model that reproduces the flux and the actual structure of the HL Tau disk.

We note that the scale height necessary to reproduce the observed asymmetry is significantly larger than the estimates from radiative transfer modeling of the 0.9-3 mm emission, H/R < 0.01 \citep{2016ApJ...816...25P}. 
One possible explanation for this inconsistency is that, in \citealt{2016ApJ...816...25P}, the focus was primarily on gaps and rings in the outer disk, rather than those within 30 au as constrained in this paper. This result then suggests that the HL Tau disk may exhibit greater vertical extent (thickness) in the inner regions while maintaining a flatter profile in the outer regions. Indeed, radial variations in scale height have been identified for the two rings of HD163296, with the 67au ring being very thick (>4au) and the 100au ring being very flat (<1au) (\citealt{2021ApJ...912..164D}, \citealt{2022SCPMA..6529511L}).

While there are limited constraints on the radial variation of scale height, this outcome implies that turbulence in disks might be higher in the inner regions compared to the outer regions. This aligns well with the high accretion rates on some disks that cannot be explained by their current low turbulence values. In this context, the magnetically driven accretion disk models suggested by \citealt{2022A&A...658A..97D} could explain the heightened turbulence parameter and increased scale height, leading to the observed asymmetry beyond the dead zone (at 23 AU) and within the MRI-active layer in the HL Tau disk. Additionally, this is consistent with what is expected in the case of vertical shear instability with dust coagulation, where the transition between thick and thin occurs at around 100 au (\citealt{2023arXiv231007332P}).
Finally, it also remains inconclusive whether the streamer observed by \citet{2022A&A...658A.104G} in HL Tau could have an effect at smaller radii through the flow of material into the inner disk. The findings in \citet{2024A&A...682A..32J} suggest a potential influence of this streamer on the turbulence at the outer radius (70 au), raising questions about a possible impact on smaller scales. Further analysis is essential to discern the factors contributing to the increased turbulence, if any, in the HL Tau inner disk and unravel the specifics of this phenomenon.

\subsection{Substructure in the inner disk: a sign of the traffic-jam effect?}

As mentioned, our highest angular resolution image at 0.45~mm revealed an increase in the brightness temperature within $\sim$2.5~au (see the inset in Figure \ref{Fig2}). There are several possible explanations for this.

First, we discard that this increase is due to free-free contamination. It is well known that HL~Tau drives a powerful jet whose emission has been previously modeled using high angular resolution VLA data at long wavelengths \citep{2019ApJ...883...71C}. Thus, we know that in the 23-43 GHz frequency range, the expected free-free emission from the radio jet is given by

\begin{eqnarray}
\left[\frac{F_{\nu}}{\mu Jy\ {beam}^{-1 }}\right ]=175\times\left [  \frac{\nu}{32.5\ GHz}\right ]^{0.7}
 \label{Eq1}
\end{eqnarray}

\noindent where $\nu$ is the frequency of the image at which we want to calculate the free-free emission. Note that in this frequency range, the free-free emission was found to be partially optically thin, which is reflected in the 0.7 spectral index. Now, since the optical depth of the free-free emission is proportional to $\nu^{-2.1}$ (e.g., \citealt{1986ApJ...304..713R}), at the frequency of our 0.45~mm image, i.e. $\sim$650~GHz, we expect free-free emission from the jet to be optically thin. Thus, there must be a turning point between 50 and 650 GHz. A conservative assumption is that emission from the jet becomes totally optically thin at 100 GHz. Then, using equation \ref{Eq1} to estimate the free-free emission at 100 GHz, and extrapolating this with a spectral index of $-$0.1 to higher frequencies, we obtain that an estimate of the free-free contamination at 650 GHz would be $\sim$300~$\mu$Jy, which is a $\sim$2\% of the peak emission at the center of the disk in the 0.45~mm image. Note that the actual contribution at this frequency should be lower since it is very likely that the free-free emission would become optically thin at a frequency lower than 100 GHz. Another potential explanation for a significant free-free emission at the central part of the disk is the possibility of it becoming highly optically thick due to photoionization. However, this scenario faces challenges because HL~Tau is a K-type star, and such stars typically do not emit sufficient ionizing photons to fully ionize a very dense gas. Taking all of these factors into account, it leads us to the conclusion that the elevated emission in the inner region of the disk is unlikely to be primarily caused by free-free contributions.

An interesting explanation for the increased dust emission at the inner part of the disk is the possibility of a dust particle pile-up occurring in that region. In fact, our unprecedented angular resolution allows us to probe the dust emissions down a few au, inside the estimated water snowline at $\sim$ 5 au if one assumes $T_{disk} = T_{B}$ (\citealp{2015ApJ...806L...7Z}, Figure \ref{Fig2}). Inside the water snowline, theoretical models predict the appearance of a traffic-jam effect (e.g \citealt{2015ApJ...815L..15B}, \citealt{2017ApJ...845...68P}), as dry silicates are more sensitive to fragmentation than ice-rich grains (\citealt{1997ApJ...480..647D}; \citealt{1997Icar..129..539S}; \citealt{2013A&A...559A..62W}; \citealt{2014acm..conf..191G}), leading to lower particle size and reduced drift speed. The traffic-jam effect offers a favored pathway for the formation of dry planetesimals in the inner disk, however, it has been recently put into question in the light of new experiments on the stickiness of ice-rich grains (e.g. \citealt{2018MNRAS.479.1273G}; \citealt{2019ApJ...873...58M}) and observations of the resilience of icy pebbles to sublimation \citep{2023arXiv231201856H}. Interestingly, \citet{Facchini2024} recently detected water emission at 321 GHz coming from the inner disk of HL Tau, but their spatial resolution was not enough to robustly measure the position of the water snowline.

Our results could be consistent with the presence of a traffic-jam effect, as such a pile-up of solids would increase the dust density and optical depth inside the water snowline (e.g. \citealt{2015ApJ...815L..15B}), causing observations to probe higher layers of the disk with more elevated dust temperatures, thereby manifesting a higher brightness temperature as we see in Figure \ref{Fig2} (See \citealt{2017A&A...608A..92D} and \citealt{2019A&A...629A..65R}). Moreover, an increase in the population of smaller particles will also result in a decrease in the albedo, which in turn results in an increase in the emission.

A second possibility is that the increase in the brightness temperature is attributed to viscous heating occurring in the inner disk. Overall, the heating mechanisms in protoplanetary disks significantly influence the temperature profile of these. The two dominant heating processes in these disks are stellar radiation and viscous heating \citep{2005ASPC..341..353D}. Across the majority of the disk's extent, stellar radiation contributes the most to the disk's thermal structure. However, in the innermost regions of the disk close to the mid-plane, where dust surface densities reach notably high values, it is highly likely that viscous heating becomes the primary contributor to the disk temperature profile \citep{2001ApJ...553..321D}.More recently, \citet{2009ApJ...705.1206C} found that the mid-plane temperature may be affected by viscous heating within the radial range of 1 to 40 au in protoplanetary disks. This phenomenon also influences the position of the snowline, which varies depending on the age of the disk. The position of these, however, is dependent upon the characteristics of the star and many other parameters which are hard to constrain.

As of now, it is challenging to distinguish between the traffic-jam or viscous heating scenarios, but we strongly encourage future works to focus on that aspect given the implications it could have concerning planet formation in the inner disk. For example, high angular resolution observations at other wavelengths could help constrain the dust emission spectral index and particle size, which would help in understanding the underlying mechanisms driving the observed changes in brightness temperature.\\

\section{Summary and conclusions}

We report new sensitive ALMA images at 0.45~mm of the HL Tau disk with an unprecedented high angular resolution of 12 mas or 1.7~au. At this wavelength, the detected emission is known to be highly optically thick. This is evidenced by the shallower substructures observed in our observation. The extended nature of the disk suggests that we are tracing smaller dust particles that are more closely coupled to the gas. 

By combining the Band 9 image with the images at other wavelengths, we have updated the model of the dust properties in the disk. The 0.45~mm image has allowed us to better constrain the dust temperature of the disk, which provided better constraints on the dust density and particle size. The particle sizes are still predicted to be larger than mm sizes until $\sim60$ au but there is a significant decrease (to $\sim$ 200 $\mu$m) between 60-100 au. These smaller particles coincide with the region where an accretion shock of infalling material has been found traced by SO and SO$_{2}$. The final derived disk dust masses are 2.1 M$_{J}$  for compact particles and 6.3 M$_{J}$ for porous dust particles. These dust masses translate into gas-to-dust ratios $\leq$ 100. 

Our 0.45~mm data clearly shows an asymmetry in the emission coming from the first ring. The emission is considerably more intense in the NE part of the disk than in the SW part. We interpreted this as a combination of very optically thick emission, a moderately inclined disk, and a large dust scale height. In the NE part of the disk, we are directly viewing the emission from the internal wall of the first ring, which is illuminated by the radiation from the central star. This interpretation is supported by our radiative transfer model which demonstrates the dependence of this asymmetry with the dust mass of the inner regions of the disk and the dust scale height. We propose that this kind of asymmetries in other disks can be used to constrain the dust scale height if an independent estimation of the dust mass of the disk has been made. Based on our proposed scenario and the multi-wavelength analysis model, we implied the possibility of a dust scale height H/R $\sim$ >= 0.08 for the HL Tau disk at least in the inner 32 au. This approximation seems in large contrast with the constraints identified by \citealt{2016ApJ...816...25P} who determined H/R < 0.01 in the outer disk. We note that this discrepancy might stem
from higher turbulence in the inner regions of HL Tau compared to its outer region leading to radial variations of the vertical structure of the HL Tau disk. Incorporating vertical structure in the multi-wavelength analysis to determine the disk mass and employing more quantitative modeling approaches, rather than qualitative ones in RADMC-3D, may be necessary to obtain a more realistic scale height of the disk.

We have resolved the innermost part of the HL Tau disk showing that there is some previously unseen substructure, revealed as a steep increase of the brightness temperature very close to the central star, at radii smaller than 5~au. Interestingly, the radius at which this increase starts appears to coincide with the position of the water ice line. Therefore, we discuss the possibility of a pile-up of small particles due to a decrease in the drift velocity of the small grains after crossing the water snowline and suffer a more efficient fragmentation. We also discuss a different possibility in which we are resolving the region where the heating of the disk changes from being dominated by irradiation to being dominated by viscous heating. Previous theoretical models predict that this change should take place very close to the protostar. To further confirm either of these scenarios, additional modeling and observations at other wavelengths are necessary for the HL Tau disk.

\begin{acknowledgements}
This paper makes use of the following ALMA data: ADS/JAO.ALMA$\#$2011.0.00005.E and ADS/JAO.ALMA$\#$2017.1.01178.S. ALMA is a partnership of ESO (representing its member states), NSF (USA) and NINS (Japan), together with NRC (Canada), MOST and ASIAA (Taiwan), and KASI (Republic of Korea), in cooperation with the Republic of Chile. The Joint ALMA Observatory is operated by ESO, AUI/NRAO and NAOJ. 

C.C.-G. acknowledges support from UNAM DGAPA-PAPIIT grant IG101321 and from CONACyT Ciencia de Frontera project ID 86372.

The research of M.V. was supported by an appointment to the NASA Postdoctoral Program at the NASA Jet Propulsion Laboratory, administered by Oak Ridge Associated Universities, under contract with NASA.
\end{acknowledgements}

%
%

\bibliographystyle{aa}
\bibliography{bibliography}
\onecolumn
\begin{appendix}

\section{MCMC model vs observations: VLA fitting}

\begin{figure*}[ht!]
\centering

\includegraphics[width=0.6\columnwidth,trim={0cm, 0cm, 0cm, 0cm},clip]{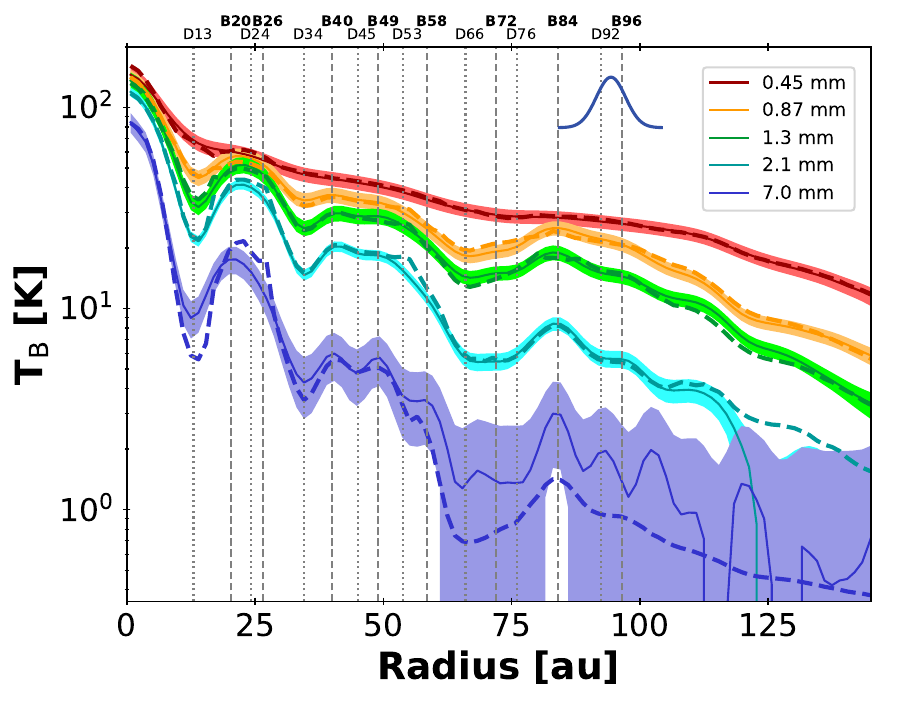}
\caption{Comparison of the brightness temperatures of the model (dash lines) with the observations (solid lines) across each radius and at each wavelength. In this Figure, the compact dust particles were used, and the VLA emission is plotted and fitted within the MCMC. The vertical lines are the positions of the gaps and bright rings taken from the dust parameters and the size of the beam is plotted as a Gaussian on the top right corner for the 5 fitted wavelengths (50 mas).} \label{Fig:A1}
\end{figure*}
\begin{figure*}[ht!]
\centering

\includegraphics[width = 0.6\columnwidth]{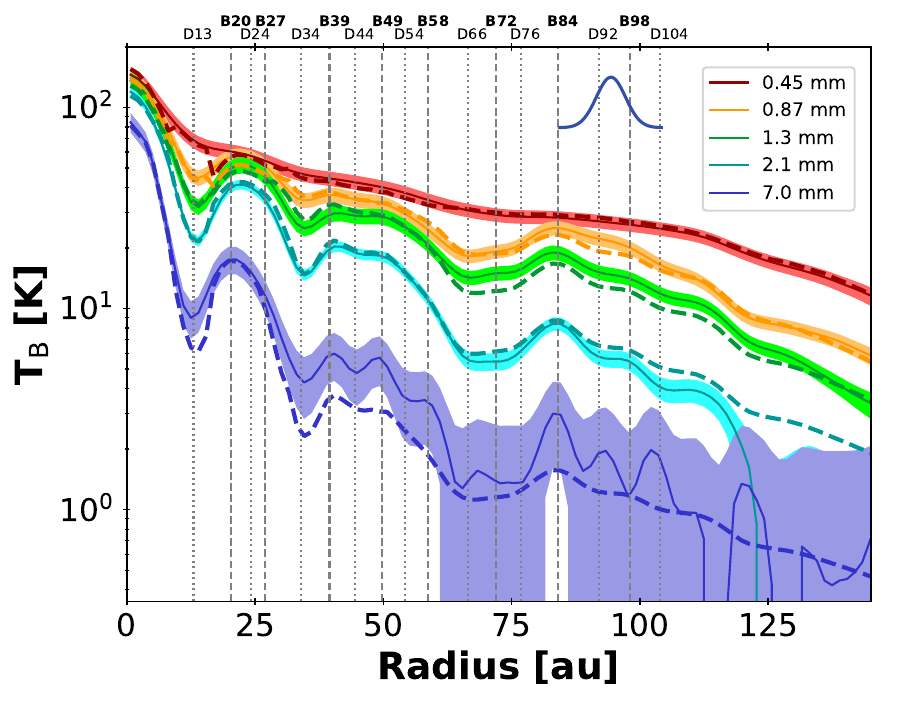}
\caption{Same as Figure B.1 but with the porous particle model instead.} \label{Fig:A2}
\end{figure*}
\end{appendix}

\end{document}